Engineering topological phase transitions via sliding ferroelectricity in $M$Bi$_2$Te$_4$ ($M$ = Ge, Sn, Pb) bilayers

Xinlong Dong[1, 2] Dan Qiao[1, 2], Zeyu Li[3], Zhenhua Qiao*[3], and Xiaohong Xu[2]*

1 College of Physics and Information Engineering, Shanxi Normal University, Taiyuan 030031, China
2 Key Laboratory of Magnetic Molecules and Magnetic Information Materials of the Ministry of Education, Research Institute of Materials Science, Shanxi Normal University, Taiyuan 030031, China
3 International Center for Quantum Design of Functional Materials, University of Science and Technology of China, Hefei 230026, China

*Corresponding author: qiao@ustc.edu.cn
*Corresponding author: xuxh@sxnu.edu.cn

Materials combining electrically switchable ferroelectricity and tunable topological states hold significant promise for advancing both fundamental quantum phenomena and innovative device architectures. Here, we employ first-principles calculations to systematically investigate the sliding ferroelectricity-mediated topological transitions in bilayer $M$Bi$_2$Te$_4$ ($M$ = Ge, Sn, Pb). By strategically engineering interlayer sliding configurations with oppositely polarized states, we demonstrate reversible band inversion accompanied by topological phase transitions. The calculated spin-orbit-coupled bandgaps reach 31 meV (GeBi$_2$Te$_4$), 36 meV (SnBi$_2$Te$_4$), and 35 meV (PbBi$_2$Te$_4$), thereby enabling room-temperature observation of the quantum spin Hall effect. Crucially, these systems exhibit substantial out-of-plane ferroelectric polarization magnitudes of 0.571-0.623 pC/m, with PbBi$_2$Te$_4$ showing the maximum polarization (0.623 pC/m). The topological nontriviality is unambiguously confirmed by two independent signatures: (i) the computed $Z_2$ topological invariant $\nu = 1$, and (ii) the emergence of gapless helical edge states spanning the bulk insulating gap. This synergy arises from the unique sliding-induced charge redistribution mechanism, which simultaneously modulates Berry curvature and breaks in-plane inversion symmetry without disrupting out-of-plane polarization stability. The co-engineering of non-volatile ferroelectric switching and topologically protected conduction channels in $M$Bi$_2$Te$_4$ bilayers establishes a material paradigm for designing reconfigurable quantum devices, where electronic topology can be electrically controlled via polarization reversal. Our results provide critical insights into manipulating correlated quantum states in van der Waals ferroelectrics for multifunctional nanoelectronics.

1 Introduction

Since the introduction of the sliding ferroelectric mechanism by researchers in 2017 [1], it has captivated significant interest within the research community due to its unique physical properties and promising application potential. The fundamental principle of sliding ferroelectricity revolves around symmetry breaking through specific interlayer stacking in bilayer and multilayer materials, which leads to the generation of reversible out-of-plane polarization [2]. The advent of sliding ferroelectricity not only established new research paradigms for ferroelectric materials but also provided innovative design principles for the development of quantum devices. Subsequent theoretical predictions and experimental investigations have further demonstrated that sliding ferroelectrics offer low switching barriers [3, 4] and high Curie temperatures [5], making them particularly appealing for high-speed, low-power consumption, and fatigue-resistant memory applications [6].

The convergence of sliding ferroelectricity with 2D material systems has unlocked unprecedented quantum phenomena. Two-dimensional crystals exhibit extraordinary property tunability spanning ferroelectricity [7-9], magnetism [10-12], quantum spin Hall (QSH) states [13-15], and quantum anomalous Hall effects [16,17]. Particularly, sliding ferroelectric coupling enables emergent functionalities including valley polarization [18-20], magnetic phase control [21-23], enhanced polarization (>1 µC/cm²) [24], and superconducting order modulation [25], while generating novel transport signatures like nonlinear Hall effects [26, 27], layer-Hall effects [28, 29], and spin-splitting phenomena [30, 31]. Despite these advances, the critical intersection between sliding ferroelectricity and QSH effects remains underexplored, a knowledge gap with profound implications for designing topological quantum devices with electrically reconfigurable edge states.

$M$Bi$_2$Te$_4$ ($M$ = Ge, Sn, Pb) systems represent a unique materials platform combining sliding ferroelectricity and topological states. Monolayers exhibit indirect bandgaps with strong spin-orbit coupling [32, 33], while bulk crystals demonstrate 3D topological insulator behavior protected by Z$_2$ invariants [34-36]. Recent theoretical advances reveal stacking-dependent topological transitions in bilayer $M$Bi$_2$Te$_4$, establishing its 2D topological insulating nature [34, 37]. Building upon these findings, we investigate the coupled sliding ferroelectric-topological physics in bilayer $M$Bi$_2$Te$_4$ through first-principles calculations.

In this study, we systematically studied the sliding ferroelectric and topological characteristics of the bilayer $M$Bi$_2$Te$_4$ ($M$ = Ge, Sn, Pb). Our results reveal that, in the $M$Bi$_2$Te$_4$ bilayer, vertically switchable ferroelectric polarization can be generated through interlayer sliding by breaking both spatial inversion symmetry and mirror symmetry. Furthermore, the topological properties of the two ferroelectric polarization states were systematically investigated, revealing nontrivial topological characteristics in both polarized configurations. The bandgaps of sliding bilayers of GeBi$_2$Te$_4$, SnBi$_2$Te$_4$, and PbBi$_2$Te$_4$ with spin-orbit coupling are 31, 36, and 35 meV, respectively. The ferroelectric polarization values for the bilayer systems of GeBi$_2$Te$_4$, SnBi$_2$Te$_4$, and PbBi$_2$Te$_4$ are 0.571, 0.615, and 0.623 pC/m, respectively. It is particularly noteworthy that the bilayer $M$Bi$_2$Te$_4$ exhibits substantial band gaps under varying ferroelectric polarization orientations, which suggests the feasibility of observing the quantum spin Hall effect at room temperature. These findings provide a theoretical foundation for the design of multifunctional quantum devices.

2 Computational methods

All calculations were performed with the framework of density functional theory as implemented in the Vienna ab initio simulation package [38, 39]. The electron-core interaction was described by the projected augmented wave [40, 41] pseudopotential with the general gradient approximation (GGA) parameterized by Perdew, Burke, and Ernzerhof (PBE) [42]. The electron wave function was expanded in plane waves up to a cutoff energy of 450 eV, and a 15 ×15 ×1 Monkhorst–Pack k-grid [43] was used to sample the Brillouin zone of the supercell. The convergence criterion for the electronic energy was set to $10^{-7}$ eV. The Hellman–Feynman force was smaller than 0.01 eV/Å in the optimized structure. To prevent interactions between neighboring layers, a vacuum space of 20 Å was established along the z-direction. Interlayer van der Waals interactions were addressed using the DFT-D3 method [44], and spin-orbit coupling (SOC) was incorporated into the electronic structure calculations. The evaluation of iron electrode electrification was performed using the Berry Phase method [45]. Ferroelectric leapfrog paths and energy barriers were computed utilizing the Nudged Elastic Band (NEB) method [46]. Moreover, the Wannier90 software package [47] was employed to construct a tight-binding (TB) model based on maximal localized Wannier functions (MLWFs), while the Wannier Tools [48] software facilitated the calculation of the Wannier charge center (WCC), $Z_2$ invariant, and boundary states to characterize the topological features of the system.

3 Results and discussion

The bulk $M$Bi$_2$Te$_4$ crystallizes in a rhombohedral layered structure (space group: R-3m), comprising alternating MTe bilayers and Bi$_2$Te$_3$ quintuple layers arranged in a Te-Bi-Te-M-Te-Bi-Te stacking sequence [49]. Each chemically bonded layer exhibits strong intra-layer covalent bonding, while adjacent layers are coupled through weak van der Waals interactions. Figure 1(a) systematically illustrates the atomic configuration of monolayer $M$Bi$_2$Te$_4$, where the top view reveals hexagonal close-packing symmetry and the side view highlights the Te-Bi-Te-M-Te-Bi-Te vertical stacking. Notably, the monolayer structure preserves in-plane inversion symmetry with the M atom serving as the central symmetric site. The lattice constants exhibit a systematic expansion with increasing M atomic radius: 4.370 Å (GeBi$_2$Te$_4$), 4.438 Å (SnBi$_2$Te$_4$), and 4.463 Å (PbBi$_2$Te$_4$), following the periodic trend of group IV elements. The preserved in-plane symmetry combined with interlayer van der Waals gaps establishes an ideal platform for investigating sliding-induced symmetry breaking phenomena.

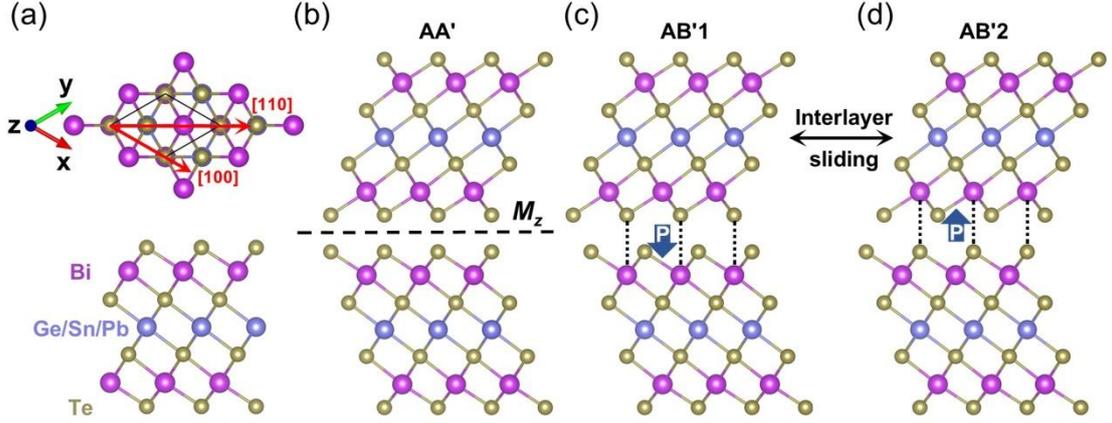

FIG. 1. (a) Atomic structure of monolayer $M$Bi$_2$Te$_4$, depicted from both top and side views. (b-d) Atomic configurations of bilayer $M$Bi$_2$Te$_4$ for (b) AA', (c) AB'1, and (d) AB'2 arrangements. The AB'1 and AB'2 structures can be formed through interlayer sliding from the AA' configuration. The blue arrows denote the direction of polarization.

Ferroelectric polarization can be introduced in bilayer $M$Bi$_2$Te$_4$ through interlayer sliding manipulation enabled by van der Waals (vdW) assembly methods. When performing a mirror operation on monolayer $M$Bi$_2$Te$_4$, the bilayer system adopts the mirror-symmetric AA' stacking configuration shown in Figure 1(b). This non-centrosymmetric bilayer structure exhibits inherent instability, where the top $M$Bi$_2$Te$_4$ layer spontaneously slides along in-plane crystallographic directions ([100] or [110], indicated by red arrows in Figure 1(a). Such interlayer sliding induces charge redistribution between layers through orbital hybridization, generating switchable ferroelectric polarization. The interlayer polarization in sliding ferroelectrics is primarily influenced by the atomic arrangements at the interface. When the top-layer Te atoms align vertically with the bottom-layer Bi atoms (Figure 1(c)), this unique non-equivalent bilayer configuration drives interfacial charge redistribution, generating out-of-plane electric polarization with downward orientation (P↓). This stacking pattern is defined as AB'1. Conversely, the AB'2 configuration emerges when top-layer Bi atoms vertically oppose bottom-layer Te atoms (Figure 1(d)), resulting in upward polarization (P↑). The interlayer stacking can be reversibly switched under external electric fields, enabling bidirectional control of polarization orientation.

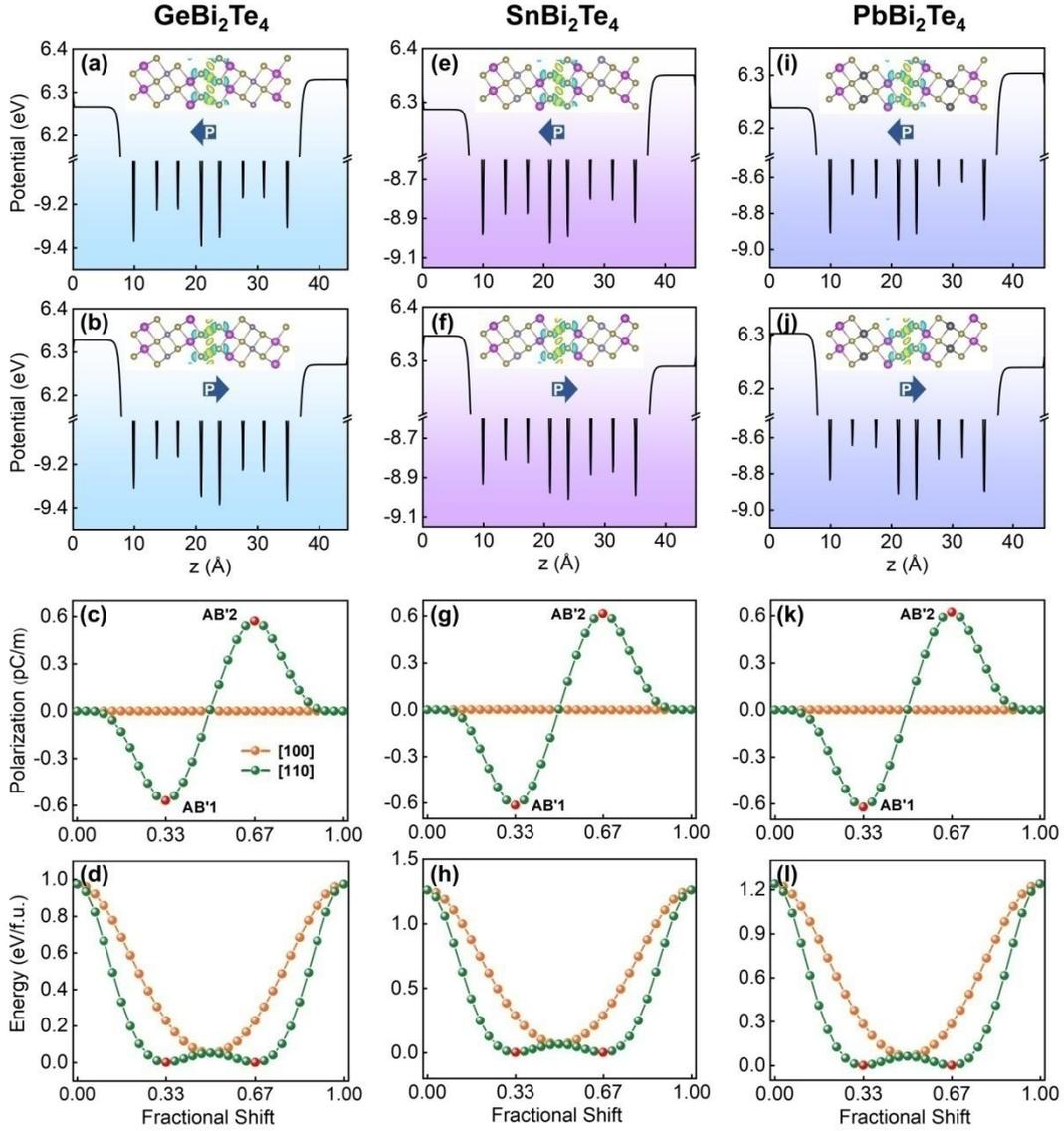

FIG. 2. Plane-averaged electrostatic potentials for the bilayer GeBi$_2$Te$_4$ in the (a) AB'1 and (b) AB'2 configurations along the z-axis. Insets in (a) and (b) illustrate the differential charge densities for the AB'1 and AB'2 arrangements, respectively, where yellow indicates electron accumulation and blue indicates electron depletion. (c) Interlayer electric polarization and (d) energy profiles related to the ferroelectric transition in bilayer GeBi$_2$Te$_4$ along the [100] and [110] directions. The AB'1 and AB'2 structures correspond to interlayer electric polarization directed downwards (P↓) and upwards (P↑), respectively, which align with the two energy minima shown as red spheres in the profiles. (e-l) correspond to SnBi$_2$Te$_4$ and PbBi$_2$Te$_4$.

To provide a clearer understanding of the polarization within the system, we analyzed the ferroelectric polarization of the bilayer GeBi$_2$Te$_4$ (see Fig 2a-d),

SnBi$_2$Te$_4$ (see Fig 2e-h), and PbBi$_2$Te$_4$ (see Fig 2i-l). Figure 2(i) illustrates the planar-averaged electrostatic potential diagram for PbBi$_2$Te$_4$ in the AB'1 stacking configuration. Here, a discontinuity value in the vacuum levels between the top (right) and bottom (left) layers ($\Delta V$ = -0.0635 eV) is observed, indicating spontaneous vertical polarization. The inset displays the differential charge density, with areas of yellow representing electron accumulation blue depicting electron depletion. This redistribution of spatial charge density reveals a depletion of electrons in the bottom layer while they accumulate in the top layer, creating a pronounced asymmetry. The reversal of the polarization sign is further evident in the plane-averaged electrostatic potential. Figures 2(i) and 2(j) distinctly illustrate opposing charge transfer directions at the interface, resulting in reversed dipole moments. For the AB'1 configuration, the ferroelectric polarization was calculated to be -0.623 pC/m using the Berry phase method.

For the AB'2 stacking configuration, it is energetically equivalent to AB'1 and can be obtained through a slip operation from AB'1. To further verify this, the planar-averaged electrostatic potential and differential charge density in the AB'2 stacking configuration were investigated. As shown in Fig. 2(k), a spontaneous vertical electric polarization in the AB'2 stacking configuration is clearly demonstrated. Consequently, the charge redistribution and $\Delta V$ in the AB'2 stacking configuration are opposite to those in the AB'1 configuration. The AB'2 stacking configuration exhibits a positive discontinuous value ($\Delta V$) of 0.0637 eV at the vacuum level, with an out-of-plane polarization value of 0.623 pC/m, corresponding to an upward vertical electric polarization. Additionally, the differential charge density reveals significant asymmetry, where electrons are depleted at the top layer and accumulate at the bottom layer. The ferroelectric polarization values for the bilayer systems of GeBi$_2$Te$_4$, SnBi$_2$Te$_4$, and PbBi$_2$Te$_4$ are 0.571, 0.615, and 0.623 pC/m, respectively, which is comparable to those of other typical sliding ferroelectrics, such as bilayer MnBi$_2$Te$_4$ (0.45 pC/m) [50, 51], bilayer MoS$_2$ (0.82 pC/m) [52], and In$_2$Se$_3$ (0.97 pC/m) [53].

TABLE 1. Positive discontinuity values ΔV, ferroelectric polarization and energy barrier for ferroelectric switching of bilayer $M$Bi$_2$Te$_4$ at P↓ and P↑.

| | | ΔV (eV) | Polarization (pC/m) | Energy (meV/f.u.) |
|---|---|---|---|---|
| GeBi$_2$Te$_4$ | P↓ | -0.0631 | -0.571 | 51.30 |
| | P↑ | 0.0575 | 0.571 | |
| SnBi$_2$Te$_4$ | P↓ | -0.0637 | -0.615 | 64.15 |
| | P↑ | 0.0569 | 0.615 | |
| PbBi$_2$Te$_4$ | P↓ | -0.0635 | -0.623 | 62.66 |
| | P↑ | 0.0637 | 0.623 | |

In addition to the opposite orientations of electric polarization, the polarization switching via interlayer sliding energy barriers is a critical factor determining the feasibility of sliding ferroelectricity. To investigate the switchability of ferroelectric polarization between the two stacking configurations, the evolution of vertical electric polarization and corresponding energy profiles during ferroelectric switching along the [100] and [110] directions were calculated using the Nudged Elastic Band (NEB) method for both AB'1 and AB'2 stacking structures, as shown in Figs. 2(c), (d), (g), (h), (k), and (l). The zero fractional displacement corresponds to the AA' stacking configuration, which exhibits high energy and instability. The ferroelectric polarization of AB'1 and AB'2 stacking configurations occurs at 1/3 (AB'1) and 2/3 (AB'2) fractional displacements along the [110] direction, where the energy is also the most stable, corresponding to downward (P↓) and upward (P↑) ferroelectric polarizations, respectively. These positions consistently represent the most stable configurations. The energy barriers for ferroelectric switching between the two stacking states in GeBi$_2$Te$_4$, SnBi$_2$Te$_4$, and PbBi$_2$Te$_4$ are 51.30, 64.15, and 62.66 meV/f.u., respectively. These values suggest that polarization switching in these materials occurs primarily through interlayer sliding, requiring only the overcoming of relatively weak interlayer van der Waals interactions. This contrasts with conventional bulk ferroelectrics, in which the switching process entails substantial distortions of atomic bonds [55]. The electrostatic potential differences (ΔV), ferroelectric polarization values, and switching energy barriers for bilayer $M$Bi$_2$Te$_4$ under P↓ and P↑ states are summarized in Table 1. These results collectively demonstrate that bilayer $M$Bi$_2$Te$_4$ materials are a promising class of sliding ferroelectric candidates.

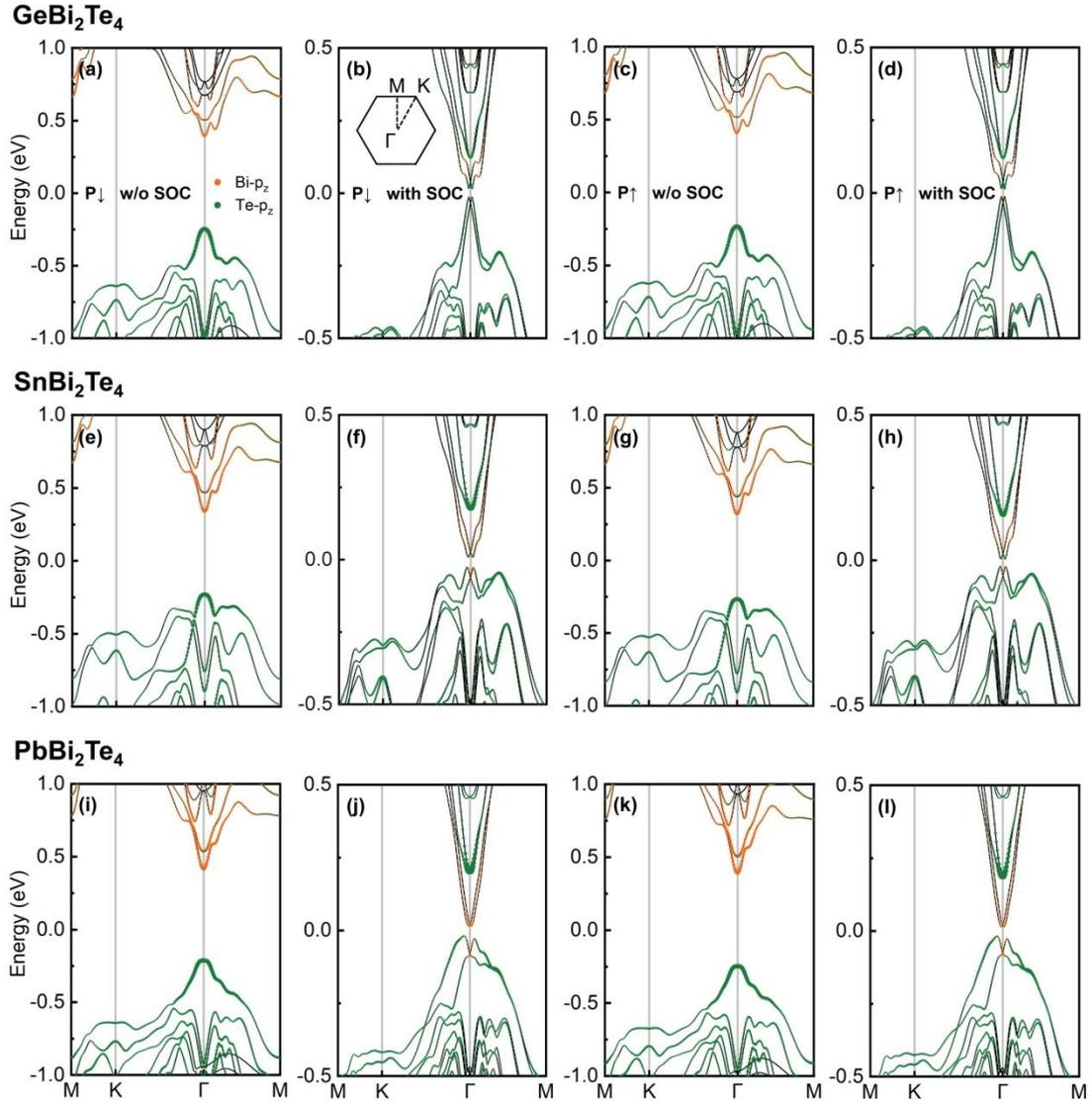

FIG. 3. Projected band structures of bilayer GeBi$_2$Te$_4$ at P↓, presented (a) without spin-orbit coupling (SOC) and (b) with SOC; (c) and (d) correspond to the P↑ configuration. The Fermi level is established at 0 eV, with the size of the points indicating the orbital contribution magnitude. (e)-l) correspond to SnBi$_2$Te$_4$ and PbBi$_2$Te$_4$.

After identifying the sliding ferroelectric potential in the system, we further investigated the band structures of bilayer $M$Bi$_2$Te$_4$ under different polarization states, as shown in Figs. 3(a)-(l). Figures 3(l) and (j) show the projected band structures of bilayer PbBi$_2$Te$_4$ in the P↓ state without spin-orbit coupling (w/o SOC) and with SOC, respectively. Without SOC, the system exhibits a global direct bandgap of 0.629 eV, where the conduction band minimum (CBM) and valence band maximum (VBM) both merge at the Γ point. When SOC is included, the bandgap reduces to 0.035 eV, and an indirect bandgap emerges across the Brillouin zone, with the CBM remaining at Γ while the VBM shifts along the K-Γ line. In both cases (with and without SOC), the electronic bands near the Fermi level are primarily contributed by the p$_z$ orbitals of the Bi and Te atoms. Under SOC, a nontrivial band inversion between Bi-pz and Te-pz orbitals suggests the presence of a nontrivial topological insulating state in the P↓ configuration of bilayer PbBi$_2$Te$_4$. Additionally, SOC induces Rashba-type spin splitting in the bands, a consequence of broken spatial inversion symmetry due to the designed interlayer vertical polarization. This symmetry breaking is further supported by the distinct electrostatic potential profiles of the two layers in Fig. 2(a), revealing an intrinsic electric field along the z-direction. Similar topological features are observed in the P↑ state of bilayer PbBi$_2$Te$_4$. The bandgap values for bilayer MBi$_2$Te$_4$ under different polarization states, with and without SOC, are summarized in Table 2. In the P↓ state, the band gaps of GeBi$_2$Te$_4$, SnBi$_2$Te$_4$, and PbBi$_2$Te$_4$, measured without spin-orbit coupling (SOC), are 0.640 eV, 0.569 eV, and 0.629 eV, respectively. When SOC is considered, these gaps diminish significantly to 0.031 eV, 0.036 eV, and 0.035 eV, respectively. In the P↑ state, the band gaps for GeBi$_2$Te$_4$, SnBi$_2$Te$_4$, and PbBi$_2$Te$_4$ are found to be 0.641 eV, 0.586 eV, and 0.638 eV, respectively. With the inclusion of SOC, the band gaps decrease further to 0.028 eV, 0.026 eV, and 0.033 eV for GeBi$_2$Te$_4$, SnBi$_2$Te$_4$, and PbBi$_2$Te$_4$, respectively. These findings highlight the interplay between sliding ferroelectricity, spin-orbit coupling, and nontrivial topology in bilayer $M$Bi$_2$Te$_4$ systems.

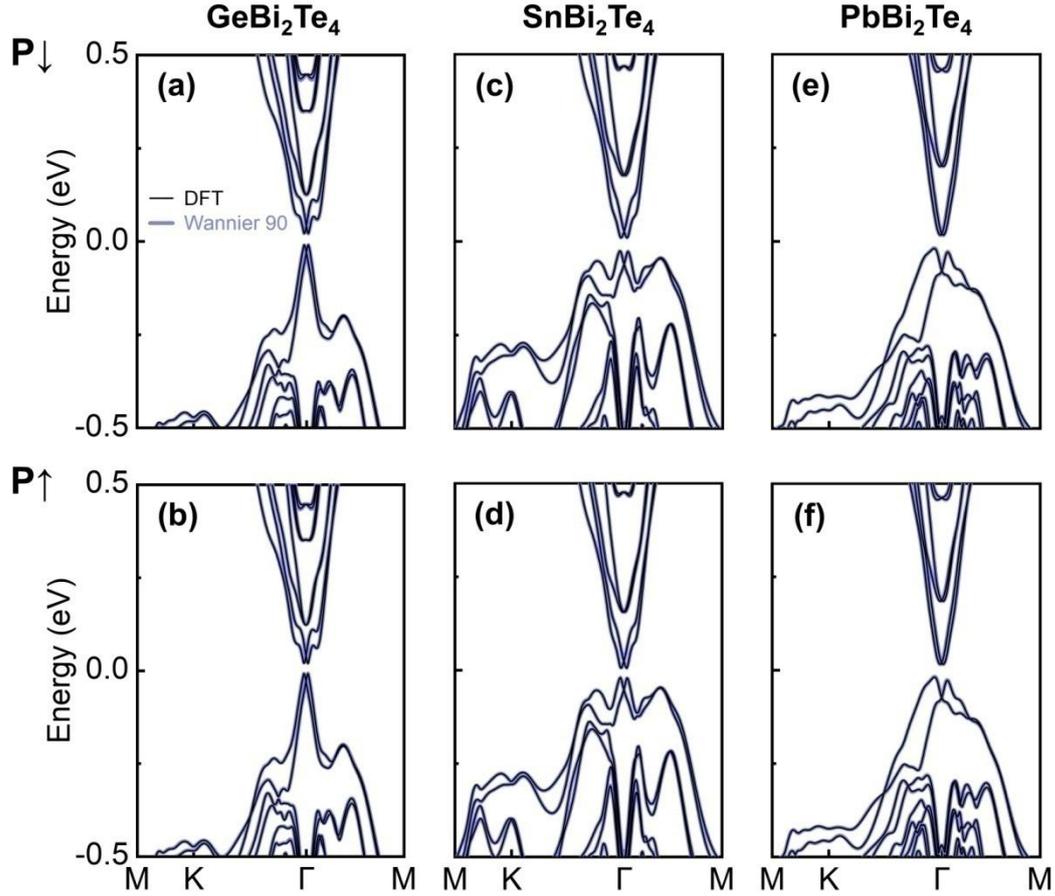

Figure 4 Band structure of bilayer $M$Bi$_2$Te$_4$ ($M$ = Ge, Sn, Pb) from Wannier function and DFT calculations. (a), (c) and (e) correspond to the P↓ configuration, (b), (d) and (f) correspond to the P↑ configuration.

To further confirm the nontrivial topological properties of bilayer $M$Bi$_2$Te$_4$, this section presents corresponding topological characterization calculations. The topological features are derived from maximally localized Wannier functions (MLWFs), constructed using the Wannier90 software package interfaced with VASP. As shown in Fig. 4(a-f), a tight-binding model based on MLWFs was established. For bilayer $M$Bi$_2$Te$_4$ under different ferroelectric polarization states, the tight-binding model successfully reproduces the DFT-calculated band structures near the Fermi level with high precision. The excellent agreement in band structures validates the reliability of the tight-binding model. Using this model, further calculations were performed with the Wannier Tools package. For two-dimensional (2D) materials with time-reversal symmetry, topological properties can be characterized by the $Z_2$ invariant. In bilayer $M$Bi$_2$Te$_4$, the introduction of interlayer ferroelectric polarization breaks spatial inversion symmetry, necessitating the use of Wannier charge centers (WCCs) to compute the $Z_2$ invariant. Figure 5(i) shows the WCC evolution for bilayer PbBi$_2$Te$_4$ in the P↓ state. Notably, the WCC trajectories exhibit no discontinuities, and any horizontal reference line intersects the WCC curves an odd number of times, indicating $Z_2 = 1$. This confirms that PbBi$_2$Te$_4$ in the P↓ state is a nontrivial topological phase. Another hallmark of 2D topological insulators is the presence of topologically protected gapless edge states. As clearly seen in Fig. 5(j), helical edge states connecting the bulk conduction and valence bands emerge within the bulk bandgap. This implies conductive edges in an otherwise insulating system, further verifying that bilayer PbBi$_2$Te$_4$ in the P↓ state is a 2D topological insulator with a nontrivial bandgap. Similar behavior is observed for the P↑ state of bilayer PbBi$_2$Te$_4$. Parallel calculations for bilayer GeBi$_2$Te$_4$ and SnBi$_2$Te$_4$, shown in Figs. 5(a–h). These findings robustly establish the interplay between sliding ferroelectricity and nontrivial topology in bilayer $M$Bi$_2$Te$_4$ materials.

TABLE 2. The bandgaps of bilayer $M$Bi$_2$Te$_4$ ($M$ = Ge, Sn, and Pb) under different polarization states.

|  |  | w/o SOC (eV) | with SOC (eV) |
| --- | --- | --- | --- |
| GeBi$_2$Te$_4$ | P↓ | 0.640 | 0.031 |
|  | P↑ | 0.641 | 0.028 |
| SnBi$_2$Te$_4$ | P↓ | 0.569 | 0.036 |
|  | P↑ | 0.586 | 0.026 |
| PbBi$_2$Te$_4$ | P↓ | 0.629 | 0.035 |
|  | P↑ | 0.638 | 0.033 |

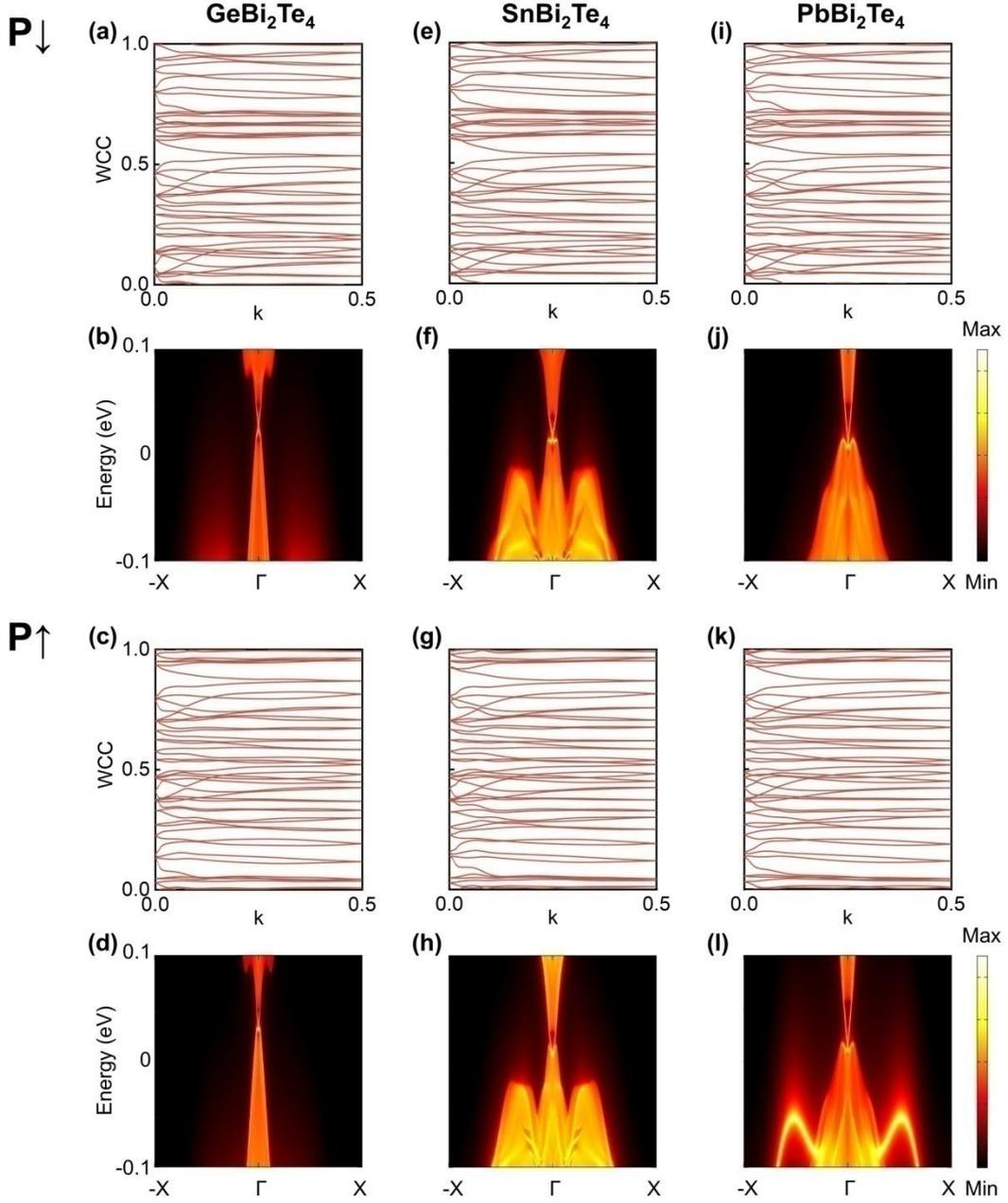

FIG. 5. (a) Evolution curves of WCCs and (b) edge states in bilayer GeBi$_2$Te$_4$ at P↓; (c) and (d) represent the configurations at P↑. (e) -l) correspond to SnBi$_2$Te$_4$ and PbBi$_2$Te$_4$.

4 Conclusions

In conclusion, our systematic investigation demonstrates that bilayer $M$Bi$_2$Te$_4$ ($M$ = Ge, Sn, Pb) exhibits electrically tunable topological states mediated by sliding ferroelectricity. First-principles calculations reveal that precise manipulation of interlayer vertical polarization enables robust realization of the quantum spin Hall effect in these heterostructures. Remarkably, The bandgaps of sliding bilayers of $GeBi_2Te_4$, $SnBi_2Te_4$, and $PbBi_2Te_4$ with spin-orbit coupling, are measured to be 31, 36, and 35 meV, respectively. These characteristics not only satisfy the prerequisite for observing quantum spin Hall states at room temperature but also establish a unique platform for co-stabilizing sliding ferroelectricity and topological insulating phases. Particularly noteworthy is the emergent sliding ferroelectric mechanism in $M$$Bi_2Te_4$ bilayers, where interlayer charge redistribution generates out-of-plane polarization that demonstrates remarkable resilience against metallic edge states. This fundamental distinction from conventional ferroelectrics ensures the preservation of ferroelectric robustness independent of topological boundary conditions. Our findings establish a paradigm for synergistic integration of non-volatile ferroelectric switching with topologically protected edge conduction in van der Waals heterostructures, opening new avenues for developing multifunctional quantum devices.

Acknowledgments

This work was supported by the National Key Research and Development Program of China (Grant No. 2022YFB3505301), the National Natural Science Foundation of China (Grant No. 52471253), the Natural Science Basic Research Program of Shanxi (Grants No. 20210302124252).

References

[1] L. Li and M. Wu, Binary compound bilayer and multilayer with vertical polarizations: two-dimensional ferroelectrics, multiferroics, and nanogenerators, ACS Nano 11, 6382 (2017)

[2] M. Wu and J. Li, Sliding ferroelectricity in 2D van der Waals materials: Related

physics and future opportunities, Proc. Natl. Acad. Sci. U.S.A. 118, e2115703118 (2021)

[3] J. Wang, X. Li, X. Ma, L. Chen, J. M. Liu, C. G. Duan, J. Íñiguez-González, D. Wu, and Y. Yang, Ultrafast Switching of Sliding Polarization and Dynamical Magnetic Field in van der Waals Bilayers Induced by Light, Phys. Rev. Lett. 133, 126801 (2024)

[4] Q. Yang and S. Meng, Light-induced complete reversal of ferroelectric polarization in sliding ferroelectrics, Phys. Rev. Lett. 133, 136902 (2024)

[5] L. Wang, J. Qi, W. Wei, M. Wu, Z. Zhang, X. Li, H. Sun, Q. Guo, M. Cao, Q. Wang, et al., Bevel-edge epitaxy of ferroelectric rhombohedral boron nitride single crystal, Nature 629, 74 (2024)

[6] P. Meng, Y. Wu, R. Bian, E. Pan, B. Dong, X. Zhao, J. Chen, L. Wu, Y. Sun, Q. Fu, et al., Sliding induced multiple polarization states in two-dimensional ferroelectrics, Nat. Commun. 13, 7696 (2022)

[7] T. Jin, J. Mao, J. Gao, C. Han, K. P. Loh, A. T. Wee, and W. Chen, Ferroelectrics-integrated two-dimensional devices toward next-generation electronics, ACS Nano 16, 13595 (2022)

[8] S. Zhou, L. You, H. Zhou, Y. Pu, Z. Gui, and J. Wang, Van der Waals layered ferroelectric $CuInP_2S_6$: Physical properties and device applications, Front. Phys. 16, 13301 (2021)

[9] F. Li, J. Fu, M. Xue, Y. Li, H. Zeng, E. Kan, T. Hu, Y. Wan, Room-temperature vertical ferroelectricity in rhenium diselenide induced by interlayer sliding, Front. Phys. 18, 53305 (2023)

[10] Y. Guo, B. Wang, X. Zhang, S. Yuan, L. Ma, and J. Wang, Magnetic two-dimensional layered crystals meet with ferromagnetic semiconductors, InfoMat. 2, 639 (2020)

[11] S. S. Wang, W. Sun, and S. Dong, Quantum spin Hall insulators and topological Rashba-splitting edge states in two-dimensional $CX_3$ ($X$= Sb, Bi), Phys. Chem. Chem. Phys. 23, 2134 (2021)

[12] X. Li, F. Liu, and Q. Wang, Na-functionalized $IrTe_2$ monolayer: Suppressed

charge ordering and electric field tuned topological phase transition, Phys. Rev. B 102, 195420 (2020)

[13] S. Zhou, C. C. Liu, J. Zhao, and Y. Yao, Monolayer group-III monochalcogenides by oxygen functionalization: a promising class of two-dimensional topological insulators, npj Quant. Mater. 3, 16 (2018)

[14] S. Pan, Z. Li, Y. Han. Electric-field-tunable topological phases in valley-polarized quantum anomalous Hall systems with inequivalent exchange fields, Front. Phys. 20, 014207 (2025)

[15] X. Zhu, Y. Chen, Z. Liu, Y. Han, Z. Qiao, Valley-polarized quantum anomalous Hall effect in van der Waals heterostructures based on monolayer jacutingaite family materials, Front. Phys. 18, 23302 (2023)

[16] X. Liu, A. P. Pyatakov, and W. Ren, Magnetoelectric coupling in multiferroic bilayer $VS_2$, Phys. Rev. Lett. 125, 247601 (2020)

[17] K. Liu, X. Ma, S. Xu, Y. Li, and M. Zhao, Tunable sliding ferroelectricity and magnetoelectric coupling in two dimensional multiferroic MnSe materials, npj Comput. Mater. 9, 16 (2023)

[18] S. Yu, Y. Xu, Y. Dai, D. Sun, B. Huang, and W. Wei, Interlayer magnetoelectric coupling in van der Waals structures, Phys. Rev. B 109, L100402 (2024)

[19] H. Li and W. Zhu, Spin-Driven Ferroelectricity in Two Dimensional Magnetic Heterostructures, Nano Lett. 23, 10651 (2023)

[20] Y. Wu, J. Tong, L. Deng, F. Luo, F. Tian, G. Qin, and X. Zhang, Coexisting ferroelectric and ferrovalley polarizations in bilayer stacked magnetic semiconductors, Nano Lett. 23, 6226 (2023)

[21] J. Ma, X. Luo, and Y. Zheng, Strain engineering the spin-valley coupling of the R-stacking sliding ferroelectric bilayer 2H-V$X_2$ (X= S, Se, Te), npj Comput. Mater. 10, 102 (2024)

[22] T. Zhang, X. Xu, B. Huang, Y. Dai, and Y. Ma, 2D spontaneous valley polarization from inversion symmetric single-layer lattices, npj Comput. Mater. 8, 64 (2022)

[23] T. Zhong, Y. Ren, Z. Zhang, J. Gao, and M. Wu, Sliding ferroelectricity in

two-dimensional Mo$A_2$N$_4$ (A= Si or Ge) bilayers: high polarizations and Moiré potentials, J. Mater. Chem. A 9, 19659 (2021)

[24] H. Wang and X. Qian, Ferroelectric nonlinear anomalous Hall effect in few-layer WTe$_2$, npj Comput. Mater. 5, 119 (2019)

[25] E. Wang, H. Zeng, W. Duan, and H. Huang, Spontaneous Inversion Symmetry Breaking and Emergence of Berry Curvature and Orbital Magnetization in Topological ZrTe$_5$ Films, Phys. Rev. Lett. 132, 266802 (2024)

[26] Y. Feng, Y. Dai, B. Huang, L. Kou, and Y. Ma, Layer Hall effect in multiferroic two-dimensional materials, Nano Lett. 23, 5367 (2023)

[27] R. Peng, T. Zhang, Z. He, Q. Wu, Y. Dai, B. Huang, and Y. Ma, Intrinsic layer-polarized anomalous Hall effect in bilayer MnBi$_2$Te$_4$, Phys. Rev. B 107, 085411 (2023)

[28] X. Chen, X. Ding, G. Gou, and X. C. Zeng, Strong Sliding Ferroelectricity and Interlayer Sliding Controllable Spintronic Effect in Two-Dimensional HgI2 Layers, Nano Lett. 24, 3089 (2024)

[29] H. Jafari, E. Barts, P. Przybysz, K. Tenzin, P. J. Kowalczyk, P. Dabrowski, and J. Sławińska, Robust Zeemantype band splitting in sliding ferroelectrics, Phys. Rev. Mater. 8, 024005 (2024)

[30] Y. Huang, H. Yuan, and H. Chen, High thermoelectric performance of two-dimensional layered $AB_2$Te$_4$ (A= Sn, Pb; B= Sb, Bi) ternary compounds, Phys. Chem. Chem. Phys. 25, 1808 (2023)

[31] R. Peng, Y. Ma, H. Wang, B. Huang, and Y. Dai, Stacking-dependent topological phase in bilayer $M$Bi$_2$Te$_4$ ($M$ = Ge, Sn, Pb), Phys. Rev. B 101, 115427 (2020)

[32] Y. Li, Y. Jia, B. Zhao, H. Bao, H. Huan, H. Weng, and Z. Yang, Stacking-layer-tuned topological phases in $M_2$Bi$_2$Te$_5$ ($M$=Ge, Sn, Pb) films, Phys. Rev. B 108, 085428 (2023)

[33] K. Yang, W. Setyawan, S. Wang, M. Buongiorno Nardelli, and S. Curtarolo, A search model for topological insulators with high-throughput robustness descriptors, Nat. Mater. 11, 614 (2012)

[34] L. Wang, Highly Tunable Band Inversion in AB$_2$X$_4$ (A=Ge, Sn,Pb, B=As, Sb, Bi;

X=Se, Te) Compounds, Phys. Rev. Materials **6**, 094201(2022)

[35] K. Okamoto, K. Kuroda, H. Miyahara, K. Miyamoto, T. Okuda, Z. Aliev, M. Babanly, I. Amiraslanov, K. Shimada, H. Namatame, M. Taniguchi, D. A. Samorokov, T. V. Menshchikova, E. V. Chulkov, and A. Kimura, Observation of a highly spin-polarized topological surface state in GeBi$_2$Te$_4$, Phys. Rev. B 86, 195304 (2012)

[36] K. Kuroda, H. Miyahara, M. Ye, S. Eremeev, Y. M. Koroteev, E. Krasovskii, E. Chulkov, S. Hiramoto, C. Moriyoshi, Y. Kuroiwa, K. Miyamoto, T. Okuda, M. Arita, K. Shimada, H. Namatame, M. Taniguchi, Y. Ueda, and A. Kimura, Experimental verification of PbBi$_2$Te$_4$ as a 3D topological insulator, Phys. Rev. Lett. 108, 206803 (2012)

[37] Y. C. Zou, Z. G. Chen, E. Zhang, F. Kong, Y. Lu, L. Wang, J. Drennan, Z. Wang, F. Xiu, K. Cho, and J. Zou, Atomic disorders in layer structured topological insulator SnBi$_2$Te$_4$ nanoplates, Nano Res. 11, 696 (2018)

[38] G. Kresse and J. Hafner, Ab initio molecular dynamics for open-shell transition metals, Phys. Rev. B 48(17), 13115 (1993)

[39] G. Kresse and J. Furthmuller, Efffciency of ab-initio total energy calculations for metals and semiconductors using a plane-wave basis set, Comput. Mater. Sci. 6(1), 15 (1996)

[40] P. E. Blöchl, Projector augmented-wave method, Phys. Rev. B 50(24), 17953 (1994)

[41] G. Kresse and D. Joubert, From ultrasoft pseudopotentials to the projector augmented-wave method, Phys. Rev. B 59(3), 1758 (1999)

[42] J. P. Perdew, K. Burke, and M. Ernzerhof, Generalized gradient approximation made simple, Phys. Rev. Lett. 77(18), 3865 (1996)

[43] H. J. Monkhorst and J. D. Pack, Special points for Brillonin-zone integrations, Phys. Rev. B 13(12), 5188 (1976)

[44] S. Grimme, J. Antony, S. Ehrlich, and H. Krieg, A consistent and accurate ab initio parametrization of density functional dispersion correction (DFT-D) for the 94 elements H-Pu, J. Chem. Phys. 132, 154104 (2010)

[45] R. D. King-Smith and D. Vanderbilt, Theory of polarization of crystalline solids, Phys. Rev. B 47, 1651 (1993)

[46] G. Henkelman, B. P. Uberuaga, and H. Jońsson, A climbing image nudged elastic band method for finding saddle points and minimum energy paths, J. Chem. Phys. 113, 9901 (2000)

[47] A. A. Mostoff, J. R. Yates, G. Pizzi, Y. S. Lee, I. Souza, D. Vanderbilt, and N. Marzari, An updated version of wannier90: A tool for obtaining maximally-localised Wannier functions, Comput. Phys. Commun. 185, 2309 (2014)

[48] Q. Wu, S. Zhang, H. F. Song, M. Troyer, and A. A. Soluyanov, WannierTools: An open-source software package for novel topological materials, Comput. Phys. Commun. 224, 405 (2018)

[49] X. Xu, D. Ni, W. Xie, and R. Cava, Superconductivity in electron-doped $PbBi_2Te_4$, Phys. Rev. B 108, 054525 (2023)

[50] R. Peng, T. Zhang, Z. He, Q. Wu, Y. Dai, B. Huang, and Y. Ma, Intrinsic layer-polarized anomalous Hall effect in bilayer $MnBi_2Te_4$, Phys. Rev. B 107, 085411 (2023)

[51] T. Cao, D. F. Shao, K. Huang, G. Gurung, and E. Y. Tsymbal, Switchable anomalous Hall effects in polarstacked 2D antiferromagnet $MnBi_2Te_4$, Nano Lett. 23, 3781 (2023)

[52] L. Gao, L. Bellaiche, Large Photoinduced Tuning of Ferroelectricity in Sliding Ferroelectrics, Phys Rev Lett, 133,196801 (2024)

[53] W. Ding, J. Zhu, Z. Wang, Y. Gao, D. Xiao, Y. Gu, Z. Zhang, and W. Zhu, Prediction of intrinsic two dimensional ferroelectrics in $In_2Se_3$ and other $III_2$-$VI_3$ van der Waals materials, Nat. Commun. 8, 14956 (2017)